\documentclass[twocolumn,superscriptaddres,prl]{revtex4}

\usepackage{graphicx}
\usepackage{epsfig}
\usepackage{amsmath, amssymb}
\usepackage{verbatim}
\usepackage{multirow}

\begin{document}

\title{
Non-Abelian Two-component Fractional Quantum Hall States
}
\author{Maissam Barkeshli}
\author{Xiao-Gang Wen}
\affiliation{Department of Physics, Massachusetts Institute of Technology,
Cambridge, MA 02139, USA }

\begin{abstract}

A large class of fractional quantum Hall (FQH) states can be
classified according to their pattern of zeros, which describes the
order of zeros in ground state wave functions as various clusters of
electrons are brought together. The pattern-of-zeros approach can be
generalized to systematically classify multilayer/spin-unpolarized FQH
states, which has led to the construction of a class of non-Abelian
multicomponent FQH states. Here we discuss some of the
simplest non-Abelian two-component states that we find and
the possibility of their experimental realization in bilayer systems
at $\nu = 2/3$, $4/5$, $4/7$, $4/9$, $1/4$, etc. 

\end{abstract}

\maketitle

There has been an ongoing effort in the condensed matter community to
experimentally realize topological phases of matter whose elementary
excitations exhibit non-Abelian statistics \cite{MR9162,W9102,NS0883}. 
While most of the attention on non-Abelian FQH states has to date been directed
towards single-component two-dimensional electron systems, 
there is good reason to look closely at two-component systems as well
(such as bilayer or spin unpolarized 
states) \cite{ES9010,EB,ME,MS9722,CY9822,SJ9147}.
Two-component quantum Hall systems allow greater variety and
tunability of effective interactions between electrons in the
partially filled Landau levels and it is the nature of these effective
interactions that ultimately determines the kind of phase that is
formed. In this letter, we report on results we have found using a
novel systematic classification of multicomponent FQH states. We will
present and discuss some of the simplest non-Abelian two-component FQH
states that we find and that occur at experimentally relevant filling
fractions.  These states may perhaps be realized in situations where
the interlayer repulsion is comparable to the intralayer repulsion. 

An important unsolved problem in FQH theory is to have a complete,
physical, and coherent understanding of how to describe the many
different FQH states that may be obtained. Such an understanding will
lead to the discovery of new topological phases of matter and, more
importantly, can give us a better overall understanding of which
non-Abelian phases are most accessible experimentally. Given the
prodigious amount of numerical and experimental effort required in
establishing the existence of a non-Abelian FQH state, it is important
to have a way to theoretically hone in on the most promising
candidates. As a step in this direction, we have constructed a
systematic classification of a large class of FQH states, which is
based on the pattern of zeros of 
wave functions. 
For example, the Laughlin wave function \cite{L8395} at $\nu = 1/m$
has an $m$th order zero as any two particles are brought together.
More generally, we can consider bringing $a$ particles together by
setting $z_i = \lambda \xi_i + z^{(a)}$ for $i = 1, ..., a$ and
expanding the wave function in powers of $\lambda$:
\begin{equation}
\Phi(\{z_i\}) = \lambda^{S_{a}} P( \{\xi_i\}; z^{(a)}, z_{a+1}, ...) 
+ O(\lambda^{S_{a} +1}).
\end{equation}
Note that the full FQH wave function is $\Psi( \{x_i, y_i \}) =
\Phi(\{z_i\}) e^{-\sum_i |z_i|^2/4l_B}$, where $z_i = x_i + i y_i$,
$\Phi(\{z_i\})$ is a polynomial in the complex coordinates $z_i$, and
$l_B$ is the magnetic length.  The sequence $\{S_a\}$ is called the
pattern of zeros and serves as a quantitative characterization of a
wide class of FQH states.  $\{S_a\}$ must satisfy certain consistency
conditions in order to describe a valid wave function $\Phi(\{z_i\})$.
Finding all valid sets of $\{S_a\}$ that satisfy these
consistency conditions then serves as a systematic classification of
FQH wave functions. Such an approach first led to a systematic
classification of non-Abelian single component quantum Hall states,
which includes the known non-Abelian states and many previously
unknown ones as well \cite{WW0808, WW0809, BW0989}.  Recently, we
have generalized the pattern-of-zeros approach to systematically
classify and quantitatively characterize non-Abelian multilayer FQH
wave functions; for a complete presentation, see \cite{BW09}.  For
$f$-component (or $f$-layer) states, the pattern of zeros is described
by a set of integers $\{S_{\vec a}\}$ indexed by a $f$-dimensional vector
$\vec a=(a_1,...,a_f)$, where $S_{\vec a}$ is the order of zeros as we
bring $a_I$ electrons together in the $I^\text{th}$ layer.  

In general, the number of integers $S_{\vec a}$ that need to be
specified is infinite in the thermodynamic limit. However, some wave
functions can be specified by much less data; the Laughlin wave
function is fully specified by $S_2$ and by the fact that there are no
off-particle zeros. The Moore-Read Pfaffian state \cite{MR9162} is
fully specified by $S_2$, $S_3$, and the fact that after combining
every pair of electrons in the Pfaffian wave function into bound
states, the induced effective wave function for the bound states
becomes a Laughlin wave function which has no off-particle zeros.
Such a $2$-cluster structure in the Pfaffian state is the reason why
$S_2$ and $S_3$ can already fully specify the state.
%
%
More generally, we believe that gapped FQH states have a $n$-cluster
structure: after combining every $n$-cluster of electrons into bound
states, the induced effective wave function for the bound states
becomes a Laughlin wave function with no off-particle zeros. 
%
%
For such $n$-cluster states, one only needs to specify $S_a$ for $a
\leq n$ to fully characterize the states. The $Z_k$ parafermion
states \cite{RR9984}, for instance, have $n = k$. The value of $n$
serves to gauge the complexity of a FQH state.  For a fixed $\nu$, as
$n$ increases, the number of topologically distinct quasiparticles,
the ground state degeneracy on higher genus surfaces and the
complexity of interactions necessary to realize the state all
increase.  This suggests that the energy gap typically decreases with
increasing $n$.  Wave functions that do not obey a cluster condition
can be thought of as having infinite $n$ and are not expected to
correspond to gapped phases.  This intuition also comes from the
conformal field theory approach to FQH wave functions; infinite $n$
corresponds to an irrational conformal field theory, which does not
yield a finite number of quasiparticles and a finite ground state
degeneracy on the torus.  In the $f$-layer case, the cluster structure
is characterized by an $f\times f$ invertible matrix: there are $f$
different kinds of clusters that can be characterized by vectors $\vec
n_I$, $I=1,...,f$.  The cluster $\vec n_I$ contains $(\vec n_I)_J$
electrons in the $J^\text{th}$ layer ($J=1,...,f$).  When all of the
electrons combine into these bound states, the resulting wave function
has a Laughlin-Halperin form with no off-particle zeros.  $S_{\vec a}$
needs to be specified only for $\vec a$ lying in the unit cell of the
lattice spanned by $\{\vec n_I\}$. In this case we may use the volume
of this unit cell as one measure of the complexity of a multilayer FQH
state and as a guide to the stability and size of the energy gap of a
FQH state. 

One of the most crucial results of the
pattern-of-zeros classification is that it gives us a broad
perspective over a large class of FQH states. So we can
determine, e.g. using the cluster structure, which states are the
simplest non-Abelian generalizations of Halperin's wave functions and
therefore which non-Abelian bilayer states are the most promising
candidates to be realized experimentally.


In the following, we will limit ourselves to describing results for
which the bilayer system is symmetric between the two layers,
which is usually (but not always) the case in experiments.  The
simplest FQH states in this case are the Halperin $(m,m,n)$ states \cite{H8375}: 
\begin{equation}
\Phi_{(m,m,n)}
= \prod_{i<j} (z_i - z_j)^m 
\prod_{i<j} (w_i - w_j)^{m} 
\prod_{i,j} (z_i - w_j)^n,
\end{equation}
which describe incompressible and Abelian FQH states at $\nu =
\frac{2}{m+n}$.  
Such a state has the simplest cluster structure described by
$(\vec n_1,\vec n_2)^T=\left(\begin{matrix} 1 & 0 \\ 0 & 1 \end{matrix} \right)$.
However these Abelian FQH states can only explain
incompressible states at $\nu = 2/p$, where $p = m+n$ is an integer.
Experiments have also seen incompressible states in two-component
systems at other filling fractions such as $\nu = 4/5$, $4/7$, $6/7$,
etc \cite{MS9722,CY9822,SJ9147}. The proposed states for these filling fractions are either two
independent single-layer phases each of which is in a hierarchy state
at $\nu = 2/5$, $2/7$, $3/7$, respectively, or some more complicated
bilayer hierarchy (e.g. composite fermion) state. If the
interlayer repulsion is comparable to the intralayer repulsion, the
existence of two independent single-layer phases is not a viable
possibility. In such a situation, it is unknown what incompressible
state would form, if any. Our pattern-of-zeros classification yields
non-Abelian states that, in addition to the bilayer composite
fermion states, should be seriously considered under these
circumstances. 

For example, we find wave functions describing non-Abelian states at
$\nu = \frac{2}{m+n}$,  at which there are also Halperin $(m,m,n)$
wave functions; the non-Abelian versions though have higher order
zeros as particles from the different layers approach each other,
indicating that they may obtain if interlayer Coulomb interactions are
comparable to intralayer interactions. We also find
interlayer-correlated non-Abelian states at $\nu = 4/p$, with $p$ odd
(e.g. 4/5, 4/7, 4/9). 
These non-Abelian FQH phases may be
more favorable than their Abelian counterparts in regimes where a
gapped bilayer FQH phase exists and where interlayer repulsion is also
strong. 

The first example that we discuss is the FQH plateau seen at $\nu =
2/3$ in bilayer systems, for which experiments have already
observed a phase transition between two FQH states \cite{LJ9792}.  The
bilayer state at this filling fraction that is usually considered
is the $(3,3,0)$ Halperin state, which consists of two independent
$1/3$ Laughlin states in each layer. Another possible bilayer state is the 
Halperin $(1,1,2)$ state, but this wave function appears somewhat unrealistic
since the order of zeros is larger when particles from different layers
approach each other than particles from the same layer. The simplest non-Abelian
bilayer states that we find appear to be more realistic; one is the
following interlayer Pfaffian state:
\begin{align}
\label{NA23}
\Psi_{2/3|_\text{inter}}
= \text{ Pf}\left(\frac{1}{x_i - x_j} \right) \Phi_{(2,2,1)} (\{z_i, w_i \}).
\end{align}
Here, $x_i$ refers to the coordinates of all of the electrons.  This
interlayer Pfaffian state may be expected if the system is
intrinsically bilayer but for which there is also strong interlayer
repulsion. Then, instead of forming the $(3,3,0)$ state, something
like the $(2,2,1)$ state would be more favorable. However the
$(2,2,1)$ state violates Fermi statistics, so we can think of adding
the Pfaffian factor in order to convert it to a valid fermion
wavefunction.  Another non-Abelian bilayer
state is the following state:
\begin{align}
\label{NA23a}
\Psi_{2/3|_\text{intra}} 
&= 
\text{ Pf}\left(\frac{1}{z_i - z_j} \right) 
\text{ Pf}\left(\frac{1}{w_i - w_j} \right) 
\Phi_{(2,2,1)} 
(\{z_i, w_i \}),
\end{align}
which has even stronger interlayer correlation.  The
$\Psi_{2/3|_\text{inter}}$ state has $2\frac12$ edge modes (\ie
central charge $c=2\frac12$) while the $\Psi_{2/3|_\text{intra}}$
state has $3$ edge modes \cite{WWH9476}. If we use the number of edge
modes to gauge the complexity of a FQH state, then the
$\Psi_{2/3|_\text{intra}}$ state is slightly more complicated than the
$\Psi_{2/3|_\text{inter}}$ state. For the cluster structure,
$\Psi_{2/3|_\text{inter}}$ has $(\vec n_1,\vec n_2)^T = \left(
\begin{matrix} 1 & 1 \\ 0 & 2 \end{matrix} \right)$ and a mimimal
charge $q_{min}=\nu/2$, while $\Psi_{2/3|_\text{intra}}$ has $(\vec n_1,\vec
n_2)^T = \left( \begin{matrix} 2 & 0 \\ 0 & 2 \end{matrix} \right)$
and a $q_{min}=\nu/4$ (see Table \ref{qpData}).  This also suggests
$\Psi_{2/3|_\text{intra}}$ to be more complicated than
$\Psi_{2/3|_\text{inter}}$.

The interlayer Pfaffian state $\Psi_{2/3|_\text{inter}}$ has in
fact been already constructed as a possible non-Abelian spin singlet
state \cite{AL0205}.  Here, we stress that, according to our systematic
classification, the non-Abelian states $\Psi_{2/3|_\text{inter}}$
and $\Psi_{2/3|_\text{intra}}$ are among the simplest of all
non-Abelian bilayer states, which indicates that they may be
experimentally viable and deserve further consideration. 

Experiments have also observed a spin-unpolarized to spin-polarized
phase transition in single-layer samples at $\nu=2/3$ \cite{ES9010}.
One candidate for the spin-unpolarized state is the $(1,1,2)$
state which has only $2$ edge modes.  However, the $(1,1,2)$ state has
very different orders of intralayer and interlayer zeros.  Thus the
spin singlet interlayer Pfaffian state $\Psi_{2/3|_\text{inter}}$ may
be more favorable than the $(1,1,2)$ state if the electron repulsion
is spin independent.  Another main candidate for the spin-unpolarized
state is a spin-singlet composite fermion state introduced in
\cite{WD9353}, which probably has the same topological order as the
$(1,1,2)$ state. 
For the single-component (or spin-polarized) phase, the candidate
states are the particle-hole conjugate of the $1/3$ Laughlin state and
the non-Abelian $Z_4$ parafermion state.

With so many different possibilities for the $\nu=2/3$ FQH state in
bilayer systems, which one is actually realized in a particular sample? Two
dimensionless quantities may be important.
The first one is $\al\equiv
V_\text{inter}/V_\text{intra}$, where $V_\text{inter}$ is the
potential for interlayer repulsion and $V_\text{intra}$ is the
potential for intralayer repulsion.  The second one is $\ga
\equiv t/V_\text{intra}$, where $t$ is the interlayer hopping
amplitude.  When $\al \sim 0$ and $\ga\sim 0$, the $(3,3,0)$ state
will be realized. If we keep $\ga\sim 0$ and increase $\al$, the
interlayer non-Abelian Pfaffian states $\Psi_{2/3|_\text{inter}}$
or $\Psi_{2/3|_\text{intra}}$ may be realized.  In the limit
$\al \sim 0$ and $\ga \gg 1$, the single-layer $\nu=2/3$ states are
realized.

A particularly interesting case is the FQH plateau observed in
two-component systems at $\nu = 4/5$. There are few proposed
explanations for two-component states at this filling fraction.  The
main proposal is that the incompressible state is described by two
independent single layer systems, each in a $2/5$-hierarchy state.
This is a reasonable possibility, considering the fact that
experiments on bilayer and wide single layer quantum wells see
incompressible states at $\nu = 2/3$, $4/5$, and $6/7$ simultaneously
\cite{MS9722}. This is twice the main sequence that one sees in single
layer samples, $1/3$, $2/5$, and $3/7$, respectively, which indicates
that perhaps each layer is forming its own independent FQH state.
However, when the interlayer repulsion between the two layers is
increased while the interlayer tunneling remains small, then the
system will undergo a phase transition into either an incompressible
state or a compressible one.

\begin{table}[t]
\begin{tabular}{|c|c|c|c|}
\hline
$\nu$ &  & Charge $q_{min}$ & Scaling Dimension $h$ \\
\hline
$2/3|_\text{inter}$ & eqn. (\ref{NA23})  & 1/3 & $\frac{1}{16} + \frac{1}{12}+0$ \\
\hline
$2/3|_\text{intra}$  & eqn. (\ref{NA23a})& 1/6 & $\frac{1}{16} + \frac{1}{48} + \frac{1}{16}$ \\
\hline
4/5 & eqn. (\ref{NA45})& 1/5 & $\frac{1}{10} + \frac{1}{40} + \frac{1}{24}$ \\
\hline
4/7 & eqn. (\ref{NA47})& 1/7 & $\frac{1}{10} + \frac{1}{56} + \frac{1}{8}$ \\
\hline
4/9 & eqn. (\ref{NA47})& 1/9 & $\frac{1}{10} + \frac{1}{72} + \frac{1}{56}$ \\
\hline
1/4 & eqn. (\ref{NA14})& 1/8 & $\frac{1}{16} + \frac{1}{32}+0$ \\
\hline
\end{tabular}
\caption{Quasiparticle minimal charges $q_{min}$ and 
the corresponding scaling dimensions $h$
for the non-Abelian bilayer states described in the given equations.
The inter-edge quasiparticle tunneling I-V curve
has a form $I \propto V^{4h-1}$ at $T=0$.
In the scaling dimension, the first term
comes from the non-Abelian part, the second term from the 
total density fluctuations (the $U(1)$ part), and the third term 
from the relative density fluctuations of the two layers (also the $U(1)$ part).
} 
\label{qpData}
\end{table}

If the system goes into a new incompressible state, then one
possibility for such a state is the following $\nu = 4/5$
non-Abelian bilayer state:
\begin{align}
\label{NA45}
\Psi( \{z_i, w_i\}) &= \Phi_{sc}( \{z_i, w_i\}) 
\Phi_{(2,2,\frac{1}{2})} (\{z_i, w_i\} ) ,
\end{align}
where $\Phi_{sc} = \langle \prod_i \psi_{1}(z_i) \psi_{2} (w_i) \rangle$ is
a correlation function in the $su(3)_2/u(1)^2$ parafermion CFT
\cite{G8710} and $\psi_1$, $\psi_2$ are Majorana fermions with scaling
dimension $1/2$. Some explicit expressions for such correlation functions
were discussed in \cite{AR0149}.  This is another one of the simplest
non-Abelian bilayer states that we find in our systematic
classification of multilayer FQH states. It is closely related to the
non-Abelian spin singlet states at $\nu = \frac{4}{4k+3}$ that were
proposed in \cite{AS9996} ($k$ is an odd or even integer for fermionic
or bosonic FQH states, respectively). 

The other major possibility for an incompressible state at $\nu = 4/5$
is that the system forms a bilayer hierarchy state, with interlayer
correlations, which would be described by a $4\times4$ $K$-matrix \cite{W9505}
and would have four edge modes. An example is the $(2/3, 2/3 | 1)$
bilayer composite fermion state \cite{SJ0113}. The primary question
then is whether it is more favorable for the system to form an Abelian
hierarchy state or a non-Abelian state.  The $su(3)_2/u(1)^2$
non-Abelian state, having only $2\frac{6}{5}$ edge modes, is simpler than the
$(2/3, 2/3 | 1)$ state. Thus, the $su(3)_2/u(1)^2$ non-Abelian
state may be more likely to appear.  
All of the states based on  $su(3)_2/u(1)^2$ have a cluster
structure $(\vec n_1,\vec n_2)^T = \left( \begin{matrix} 2 & 0 \\ 0 &
2 \end{matrix} \right)$, and a mimimal charge $q_{min}=\nu/4$ (see
Table \ref{qpData}).



Similar discussions hold also for FQH states at $\nu = 4/7$ and $\nu
= 4/9$.  An incompressible state has been observed at $\nu = 4/7$ in
wide quantum wells \cite{MS9722}, but not to our knowledge at $\nu  =
4/9$. On the other hand, phase transitions have been observed at these
filling fractions in single layer systems, purportedly between
spin-polarized and spin-unpolarized states \cite{CY9822}.  This
suggests an incompressible state at $\nu = 4/9$ may also be observed
in bilayer or wide single layer quantum wells if the system can
be made clean enough and the interlayer repulsion be made comparable
to the intralayer repulsion while keeping the interlayer tunneling
small. Among the simplest non-Abelian bilayer states that we find
through the pattern-of-zeros classification is the non-Abelian
spin-singlet state at $\nu = 4/7$, which was already proposed in
\cite{AS9996}, and a close relative at $\nu = 4/9$:
\begin{align}
\label{NA47}
\Phi_{sc}( \{z_i, w_i\}) 
\Phi_{(2,2,\frac{3}{2})} (\{z_i, w_i\} ) & \;\; \nu = 4/7,
\nonumber \\
\Phi_{sc}( \{z_i, w_i\}) 
\Phi_{(4,4,\frac{1}{2})} (\{z_i, w_i\} ) & \;\; \nu = 4/9.
\end{align}
As before, $\Phi_{sc} = \langle \prod_i \psi_{1}(z_i) \psi_{2} (w_i)
\rangle$ is a correlation function in the $su(3)_2/u(1)^2$
parafermion CFT.

Recently, an incompressible state was found at $\nu = 1/4$ and it is
unclear what phase this corresponds to and even whether it is a
single-layer or bilayer phase \cite{LP0874}. Some possibilities that have
recently been considered \cite{PM0915} are the $(5,5,3)$ and $(7,7,1)$
Halperin states and the $\nu = 1/4$ single-layer Pfaffian. The pattern
of zeros construction yields many other alternative possibilities,
perhaps the most physical (and simplest) of which is the following
interlayer Pfaffian:
\begin{equation}
\label{NA14}
\Psi(\{z_i, w_i \}) = \text{ Pf}\left(\frac{1}{x_i - x_j} \right) \Phi_{(6,6,2)} (\{z_i, w_i \}).
\end{equation} 


A useful tool for identifying FQH states in numerical studies of exact
diagonalization on finite systems on a sphere is to look at what values of the
shift, $\mathcal{S} = \nu^{-1} N_e - N_{\Phi}$, a ground state with
zero total angular momentum is found \cite{WZ9253}. This then limits
the possibilities of which topological phase is realized in the system
to those that have that particular value of the shift. Similarly, in
numerical studies of multilayer systems, one can look for the
different sets $(N_1,\cdots, N_f; N_{\Phi}^1,\cdots, N_{\Phi}^{f})$
that yield a ground state with zero total angular momentum. Here
$N_I$, $N_{\Phi}^I$ are the number of particles and number of flux
quanta, respectively, in the $I$th layer ($I = 1, ..., f$).
Each topological phase will have its own list of
$(N_1,\cdots, N_f; N_{\Phi}^1,\cdots, N_{\Phi}^{N_f})$ that let it
fill the sphere; analyzing this can be a useful way of determining
which topological phase is obtained numerically. In \cite{BW09}, we
have found conditions that $\vec N$ and $\vec N_{\Phi}$ should satisfy
for the FQH state to fill the sphere. 
For the states presented here, $N_1$ and $N_2$ must be even, and they
determine $N_{\Phi}^1$, $N_{\Phi}^2$ through:
$ \left( \begin{matrix} N_{\Phi}^1 + \mathcal{S} \\
N_{\Phi}^2 + \mathcal{S} \end{matrix} \right) = M \left(\begin{matrix}
N_1 \\ N_2 \end{matrix} \right), $ where $\mathcal{S}$ is the shift 
on the sphere and $M$ is a matrix. For these states, which are of the form
$\Phi = \Phi_{sc} \Phi_{(m,m,l)}$, 
$M = \left( \begin{matrix} m & l \\ l & m \end{matrix} \right) $.  
The value of the shifts are listed in Table
\ref{proposed}.

This research is partially supported by NSF Grant No.  DMR-0706078.


\begin{center}
\begin{table}
\begin{tabular}{|c|l|c|c|}
\hline
 $\nu$ &  Proposed States & Edge Modes & Shift $\mathcal{S}$ \\
\hline
\hline
\multirow{6}{*}{$2/3$} & $(3,3,0)$ & $2$ & $3$ \\
      & $(1,1,2)$ & $2$ &  $1$ \\  
      & $2/3|_\text{inter}$ (see eqn. (\ref{NA23}))& $2\frac12$ & $3$ \\
      & $2/3|_\text{intra}$ (see eqn. (\ref{NA23a}))& $3$ & $3$ \\
      & $Z_4$ parafermion & $3$ & 3 \\
      & P-H conjugate of $\nu=1/3$ & $1_R+1_L$ & 0 \\
\hline
\multirow{2}{*}{$4/5$} & $(2/5, 2/5 | 0)$ & 4 & 4 \\
& $su(3)_2/u(1)^2$ (see eqn. (\ref{NA45})) & $2\frac{6}{5}$ & 3 \\
& $(2/3, 2/3 | 1)$  & $2_R+2_L $  & 0 \\
\hline
\multirow{4}{*}{$4/7$} & $(2/7,2/7|0)$ & 4 & 2 \\     
& $su(3)_2/u(1)^2$ (see eqn. (\ref{NA47})) & $2\frac{6}{5}$ & 3 \\
& $(2/5, 2/5 | 1)$ & 4 & 4 \\
& $(2/3, 2/3 | 2)$ & $1_R+3_L $ & 0 \\
\hline
\multirow{4}{*}{$1/4$} & $(5,5,3)$ & 2 & 5 \\
      & $(7,7,1)$ & 2 & 7 \\
      & Inter-layer Pfaffian (see eqn. (\ref{NA14})) & $2\frac12$ & 7 \\
      & Single-layer Pfaffian & $1\frac12$ & 5 \\
\hline
\end{tabular}
\caption{ Proposed explanations for incompressible states at experimentally relevant 
filling fractions, $\nu = 2/3$, $4/5$, $4/7$, and $1/4$,
in two-component FQH systems. The bilayer composite fermion state $(\nu_1, \nu_2 | m)$ \cite{SJ0113}
refers to the state $\prod_{i,j} (z_i - w_j)^m \Phi_{\nu_1}( \{z_i\}) \Phi_{\nu_2} (\{w_i\})$, where $\Phi_{\nu}$
is a single layer composite fermion state at filling fraction $\nu$. For $(2/3,2/3|m)$, we have
taken the single layer $2/3$ state to be the particle-hole conjugate of the Laughlin state. $n_R + n_L$ 
indicates that there are $n_R$ right-moving edge modes and $n_L$ left-moving edge modes. 
}
\label{proposed}
\end{table}
\end{center}

\end{document}